\documentclass[11pt]{article}
\usepackage{moriond,epsfig,psfig}

\def\be{\begin{equation}}
\def\ee{\end{equation}}
\def\bea{\begin{eqnarray}}
\def\eea{\end{eqnarray}}

\begin{document}
\hfill{IFIC/05-24}
\vspace*{4cm}
\title{MASSIVE NEUTRINOS AND COSMOLOGY}

\author{SERGIO PASTOR}

\address{Instituto de F\'{\i}sica Corpuscular (CSIC-Universitat de
Val\`encia)\\ Ed.\ Institutos de Investigaci\'on, Apdo.\ 22085,
E-46071 Valencia, Spain}

\maketitle\abstracts{Cosmology can provide information on the absolute
  scale of neutrino masses, complementary to the results of tritium
  beta decay and neutrinoless double beta decay experiments. We show
  how the analysis of data from the anisotropies of the cosmic
  microwave background radiation and from the distribution of
  cosmological large-scale structure, combined with other experimental
  results, provides an upper bound on the sum of neutrino masses. We
  also discuss the sensitivity of future cosmological data to neutrino
  masses.}

\section{Introduction}

Neutrinos are very abundant in the universe, in number only slightly
smaller than that of relic photons. After being created in earlier
epochs, relic neutrinos influence various cosmological stages, playing
an important role that has been used to derive bounds on non-standard
neutrino properties, alternative to the limits from terrestrial
neutrino experiments, in some cases the only available. For an
extensive review on many aspects of neutrino cosmology, see e.g.\
\cite{Dolgov:2002wy}.

In this contribution we focus on the connection between cosmology and
neutrino masses, reviewing recent cosmological bounds on the sum of
neutrino masses, in particular those that appeared after the release
of the first year data of the Wilkinson Microwave Anisotropy Probe
(WMAP) \cite{Bennett:2003bz}. We also discuss how future cosmological
observations will improve the sensitivity to neutrino masses in the
sub-eV region.

\section{The cosmic neutrino background}

In the early universe neutrinos were in thermal equilibrium through
the standard weak interactions with other particles. Thus the
distribution of neutrino momenta was a Fermi-Dirac one,
\begin{equation}
f_{\nu_\alpha}(p)=\left [\exp \left
(\frac{p-\mu_{\nu_\alpha}}{T}\right)+1 \right]^{-1}~, 
\label{FDspectrum}
\end{equation}
for neutrino masses much smaller than the temperature. The chemical
potentials of the different neutrino flavours $\mu_{\nu_\alpha}$ could
have been non-zero if an asymmetry between the number of neutrinos and
antineutrinos was previously created. Although this lepton asymmetry
can still be very large compared to the baryon one, it has been shown
\cite{Dolgov:2002ab} that in practice due to the effectiveness of
flavour neutrino oscillations before the onset of primordial
nucleosynthesis, it can be safely neglected.

As the universe cools, at a temperature $T_{\rm dec}\simeq {\cal
O}$(MeV) the weak interaction rate $\Gamma_\nu$ falls below the
expansion rate given by the Hubble parameter $H$ and neutrinos
decouple from the rest of the plasma. After decoupling, the
collisionless neutrinos expand freely, keeping a phase-space density
corresponding to that of a relativistic species in equilibrium, and do
not essentially share the entropy transfer from $e^+e^-$ annihilations
into photons that causes the well-known temperature difference
$T_\gamma/T_\nu = \left( 11/4 \right)^{1/3}$ between relic photons and
relic neutrinos
\footnote{Some residual interactions between $e^+e^-$ and neutrinos
lead to small distortions on the neutrino spectra with respect to that
in Eq.\ \ref{FDspectrum}. The effect over the relativistic degrees of
freedom corresponds to $N_{\rm eff}=3.045$ (see
\cite{Mangano:2001iu,Tegua} for the latest analyses and previous
references).}. It is thus easy to calculate the number and energy
densities of relic neutrinos at later epochs. The former is fixed by
the value of the temperature (approximately there are now 112
neutrinos and antineutrinos per flavour and cm$^{-3}$), but the energy
density is a function of the mass that should be in principle
calculated numerically, with the analytical limits
\begin{eqnarray}
\rho_\nu (m_\nu \ll T_\nu)&=& 
\frac{7\pi^2}{120}
\left(\frac{4}{11}\right)^{4/3}\;T_\gamma^4
\nonumber \\
\rho_\nu (m_\nu \gg T_\nu)&=&  \sum_i  m_i \, n_\nu~\
\label{nurho}
\end{eqnarray}
where the sum runs over all neutrino states for which $m_i \gg T_\nu$.
For values of neutrino masses much larger than the present cosmic
temperature ($T_\gamma\sim T_\nu\approx 10^{-4}$ eV), one finds that
the contribution of neutrinos to the total energy density of the
universe is, in terms of its critical value $\rho_c$,
\begin{equation}
\Omega_\nu = \frac{\rho_\nu}{\rho_c} = \frac{\sum_{i} m_i}
{93.2\,h^2~{\rm eV}}
\label{Omeganu}
\end{equation}
where $h\equiv H_0/(100$ km s$^{-1}$ Mpc$^{-1})$ is the present value
of the Hubble parameter.

\section{Neutrinos as dark matter}

%Basically neglect sterile

The role of neutrinos as dark matter (DM) particles has been widely
discussed since the 1970s. Two facts favour massive neutrinos as DM:
they definitely exist and it is enough to have eV masses in order to
produce a contribution of order unity to the present energy density of
the universe. {}From Eq.\ \ref{Omeganu} one easily finds an upper
limit on the masses (some tens of eV) by imposing the very
conservative bound $\Omega_\nu<1$.

The background of relic massive neutrinos affects the evolution of
cosmological perturbations in a particular way: it erases the density
contrasts on wavelengths smaller than a mass-dependent free-streaming
scale. This damping of the density fluctuations on small scales is
characteristic of hot dark matter (HDM) particles. In a universe
dominated by HDM, large objects such as superclusters of galaxies form
first, while smaller structures like clusters and galaxies form via a
fragmentation process (a top-down scenario). However, within the
presently favoured $\Lambda$CDM model, dominated at late times by dark
energy and where the main matter component is pressureless, there is
no need for a significant contribution of HDM. Therefore, one can use
the available cosmological data to find how large the neutrino
contribution can be. If all neutrino states have the same spectrum, as
in the standard cosmological model, an analysis of the data will
provide an upper bound on the sum of all neutrino masses.

This bound is important because presently we have experimental
evidences of flavour neutrino oscillations, which are sensitive to the
squared mass differences between the three neutrino mass states
$m_{1,2,3}$: allowed $3\sigma$ ranges are $\Delta
m_{23}^2=(1.4-3.3)\times 10^{-3}$ eV$^2$ and $\Delta
m_{12}^2=(7.2-9.1)\times 10^{-5}$ eV$^2$ (see 
e.g.\ \cite{lisi,Maltoni:2004ei} and references therein).  These values are
perfectly compatible with a hierarchical scenario where the neutrino
massive states have $m_1\sim 0$, $m_2\sim (\Delta m_{12}^2)^{1/2}$ and
$m_3\sim (\Delta m_{23}^2)^{1/2}$ (or with an inverted hierarchy where
$m_3 \sim m_2 \sim (\Delta m_{23}^2)^{1/2}$, separated by the small
$\Delta m_{12}^2$, see e.g.\ Fig.\ 1 in \cite{Lesgourgues:2004ps}).
The sum of neutrino masses would then be of the order $\sum_i m_i
\simeq m_3 \sim 0.05$ eV (or $\sum_i m_i \simeq m_3+m_2 \sim 0.1$ eV
in the inverted case).  Alternatively, the three states could be
degenerate, with masses much larger than the differences, so that
$\sum_i m_i \simeq 3m_0$.

Cosmology is at first order sensitive to the total mass if all
neutrino states have the same number density, providing information on
$m_0$ but blind to neutrino mixing angles or possible CP violating
phases.  This fact differentiates cosmology from terrestrial
experiments such as beta decay and neutrinoless double
beta decay \cite{2nubb}, which are sensitive to $\sum_i | U_{ei}|^2
m_i^2$ and $m_{ee}\equiv |\sum_i U_{ei}^2 m_i|$, respectively ($U$ is
the $3\times 3$ mixing matrix that relates the weak and mass
bases). Presently, from tritium beta decay one finds $m_0<2.2$ eV ($95
\%$ CL), a bound expected to be improved by the KATRIN project to
reach $0.3-0.35$ eV \cite{katrin}. We also have results on $m_{ee}$
from neutrinoless double beta decay experiments, which give upper
bounds in the range $0.3-1.6$ eV and a claim of positive evidence for
$m_{ee}$ \cite{Klapdor-Kleingrothaus:2004wj}. However, these results
suffer from the uncertainties in the calculations of the corresponding
nuclear matrix elements (for a review, see e.g.\ \cite{2nubb_rev}).

\section{Current cosmological bounds on neutrino masses}

For neutrino masses of order eV, the free-streaming effect can be
detectable in the linear matter power spectrum, reconstructed from
galaxy redshift surveys (a rough analytical approximation of the effect
is $\Delta P(k)/P(k) \sim -8 \Omega_\nu/\Omega_m$ \cite{Hu:1997mj},
usually quoted in the literature). Massive neutrinos have also a
smaller background effect: different values of the neutrino density
fraction $\Omega_{\nu}$ have to be compensated by small changes to the
other components, modifying some characteristic times and scales in
the history of the universe, like the time of equality between matter
and radiation, or the size of the Hubble radius at photon
decoupling. Although neutrino masses influence only slightly the
spectrum of the anisotropies of Cosmic Microwave Background (CMB)
radiation, it is crucial to combine CMB and large-scale structure
(LSS) observations, as well as other cosmological observations, in
order to measure the neutrino mass, because CMB data give independent
constraints on the cosmological parameters, and partially removes the
parameter degeneracies.

\begin{table}[t]
\caption{Upper bounds on $\sum m_\nu$ from recent
analyses of different sets of cosmological data.\label{table:1}}
\vspace{0.4cm}
\begin{center}
\begin{tabular}{|c|c|l|}
\hline
{}&{}&{}\\
Ref.\ & Bound on  $\sum m_\nu$ (eV, 95\% CL)& Data (in addition to WMAP)\\
{}&{}&{}\\
\hline
{}&{}&{}\\
\cite{Ichikawa:2004zi}     & 2.0 & --\\
{}&{}&{}\\
\cite{Tegmark:2003ud}     & 1.7 & SDSS\\
{}&{}&{}\\
\cite{Hannestad:2003xv}   & 1.0 & other CMB, 2dF, HST, SN\\
{}&{}&{}\\
\cite{Crotty:2004gm}      & 1.0 [0.6] & other CMB, 2dF, SDSS [HST, SN]\\
{}&{}&{}\\
\cite{Barger:2003vs}      & 0.75 & other CMB, 2dF, SDSS, HST\\
{}&{}&{}\\
\cite{Spergel:2003cb}     & 0.7 & other CMB, 2dF, $\sigma_8$, HST\\
{}&{}&{}\\
\cite{Fogli:2004as}      & 0.47 & other CMB, 2dF, SDSS (Ly-$\alpha$),
HST, SN\\{}&{}&{}\\
\cite{Seljak:2004xh}      & 0.42 & SDSS (bias, galaxy clustering, Ly-$\alpha$)\\
{}&{}&{}\\
\hline
\end{tabular}
\end{center}
\end{table}

We show in Table \ref{table:1} a summary of recent results from
analyses of cosmological data, which emphasizes the fact that a single
cosmological bound on neutrino masses does not exist. Assuming that
the relic neutrinos are standard, the limits depend on the underlying
model (the set of cosmological parameters) and the cosmological data
used. The data include CMB experiments (WMAP, other CMB such as ACBAR
and CBI) and different LSS data: the distribution of galaxies from
2dFGRS or the Sloan Digital Sky Survey (SDSS), the {\em bias}
(normalization of the matter power spectrum, for instance through a
parameter such as $\sigma_8$) or the matter power spectrum on small
scales inferred from the Lyman-$\alpha$ forest. In addition, other
cosmological data can be incorporated via priors on parameters such as
$h$ (HST) or $\Omega_m$ (SNI-a data).  For details, we refer the
reader to the discussion in \cite{Crotty:2004gm}.

%non-linear corrections to $P(k)$
It is important to emphasize that not all the analyses in Table
\ref{table:1} used the same set of cosmological parameters and priors
(such as the assumption of a flat Universe), so that minor
differences on the quoted bounds exist even when using the same
cosmological data. Actually there exist some works where non-zero
neutrino masses could be either favoured by cosmology if radical
departures from the {\em standard} $\Lambda$CDM model are considered
(see e.g.\ \cite{Blanchard:2003du}) or needed to fit some particular
data, as in \cite{Allen:2003pt} where the region of allowed neutrino
masses was found to be $\sum m_\nu=0.56^{+0.30}_{-0.26}$ eV when using
a a low value of the normalization of the matter power spectrum from
X-ray cluster data. These examples show that the sensitivity of
cosmological observations to neutrino masses is a powerful tool, but
its implications should not be extracted without care.

One can see from Table \ref{table:1} that a conservative cosmological
bound on $\sum m_\nu$ of the order 1 eV was found from CMB results
combined only with galaxy clustering data from 2dFGRS and/or SDSS
(i.e.\ the shape of the matter power spectrum for the relevant
scales). The addition of further data via priors improves the bounds,
which reach the lowest values when Lyman-$\alpha$ data (from SDSS) are
included as in refs.\ \cite{Seljak:2004xh,Fogli:2004as}: the
contribution of a total neutrino mass of the order $0.4$-$0.5$ eV seems
already disfavoured.

\begin{figure}[t]
%\hspace{0.225\textwidth}
%\begin{center}
  \epsfig{figure=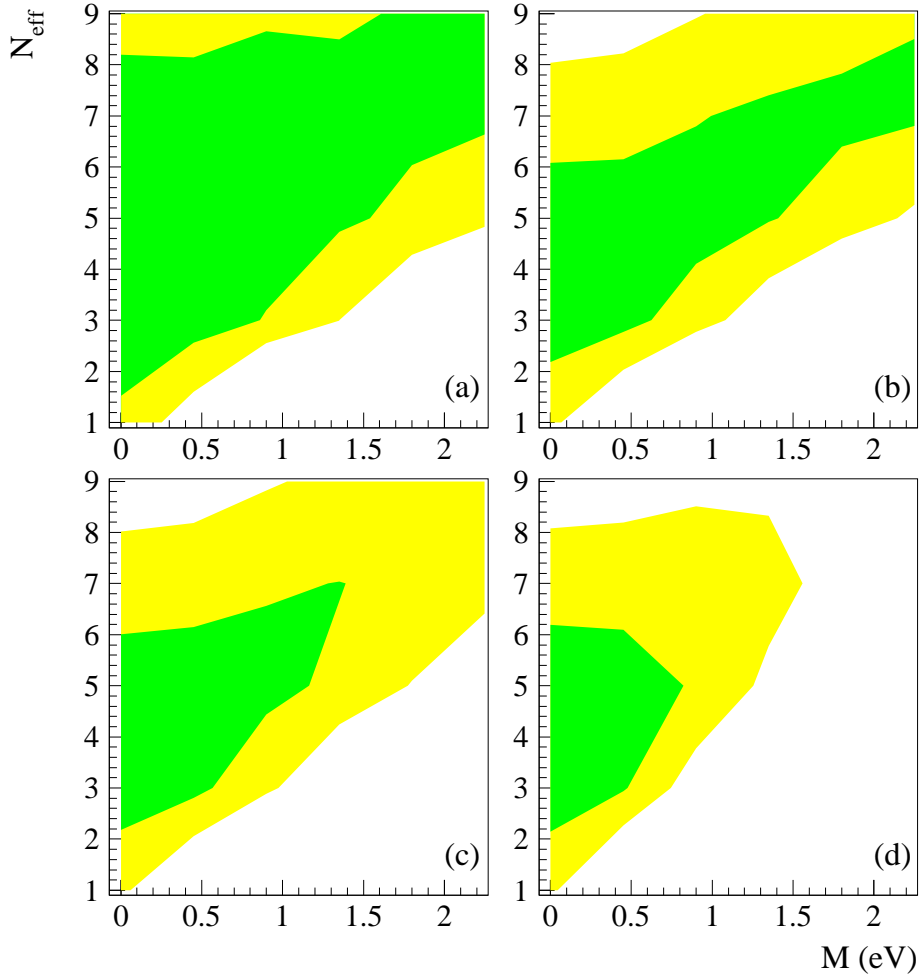,width=0.85\textwidth}
%\end{center}
\caption{\label{fig:Nandm} Two-dimensional likelihood in $(N_{\rm
eff}, M\equiv \sum m_\nu)$ space, marginalized over the other
cosmological parameters of the model. We plot the 1$\sigma$ (green /
dark) and 2$\sigma$ (yellow / light) allowed regions. Here we used CMB
(WMAP \& ACBAR) and LSS (2dF \& SDSS) data, adding extra priors on $h$
(HST) and $\Omega_m$ (SN99 or SN03) as follows: (a) no priors, (b)
HST, (c) HST+SN99, (d) HST+SN03. For details, see ref.\
\protect\cite{Crotty:2004gm}.}
\end{figure}

An interesting case of degeneracy between cosmological parameters is
that between neutrino masses and the radiation content of the universe
(parametrized via the effective number of neutrinos $N_{\rm
eff}$). The extra radiation partially compensates the effect of
neutrino masses, leading to a less stringent bound on $\sum m_\nu$
\cite{Hannestad:2003xv,Crotty:2004gm,Hannestad:2003ye}, as shown in
Fig.\ \ref{fig:Nandm}. This applies to the neutrino mass schemes that
also explain the results of LSND, an independent evidence of neutrino
conversions at a larger mass difference than those quoted in the
previous section, where a fourth sterile neutrino is required with
mass of ${\cal O}$(eV) \cite{Maltoni:2004ei}.  At present, the LSND
regions in the space of oscillation parameters (that will be checked
by the ongoing MiniBoone experiment \cite{MiniBoone} are not yet
completely disfavoured by cosmological data.

Finally, let us remind the reader that the cosmological implications
of neutrino masses could be very different if the spectrum or
evolution of the cosmic neutrino background is non-standard. For
instance, the bounds on neutrino masses would be modified if relic
neutrinos have non-thermal spectra \cite{Cuoco:2005qr} or violate the
spin-statistics theorem obeying Bose statistics \cite{Dolgov:2005mi},
or could almost completely disappear, such as for the case of mass
varying neutrinos (see e.g.\ \cite{Fardon:2003eh} and references
therein).

\section{Future sensitivities to neutrino masses from cosmological 
observations}

\begin{figure}[t]
%\hspace{0.225\textwidth}
\begin{center}
  \epsfig{figure=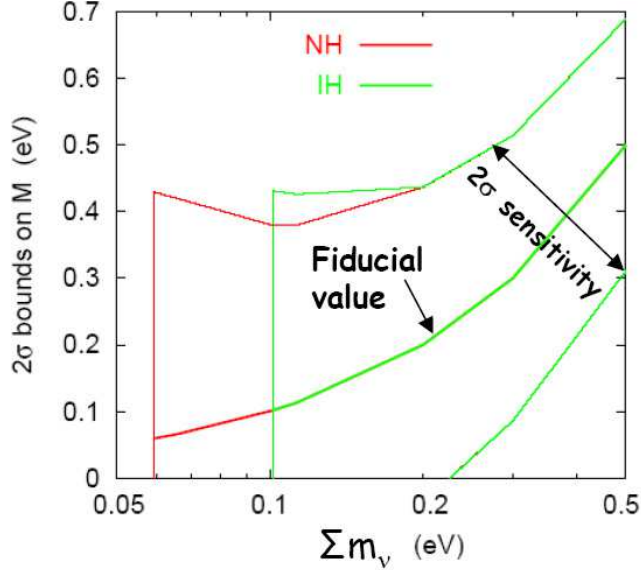,width=0.55\textwidth}
\end{center}
\caption{\label{fig:Planck} Predicted 2$\sigma$ error on the total
neutrino mass $M \equiv \sum m_\nu$ as a function of $M$ in the
fiducial model, using future data from PLANCK and SDSS (limited to
$k_{\rm max}=0.15~h$ Mpc$^{-1}$). The case NH (IH) corresponds to
three massive neutrino states where the total neutrino mass is
distributed according to a normal (inverted) hierarchy.  For details
of the analysis, see ref.\ \protect\cite{Lesgourgues:2004ps}.}
\end{figure}

Future CMB data from WMAP and Planck, combined with LSS data from
larger galaxy surveys will enhance the cosmological sensitivity to
neutrino masses. The pioneering calculation in ref.\ \cite{Hu:1997mj}
found that the combination of Planck and SDSS data will push the bound
on $\sum m_\nu$ to approximately $0.3$ eV at $95\%$ CL. An updated
forecast analysis \cite{Hannestad:2002cn} lowered this value to $0.12$
eV, almost reaching the values in the hierarchical scenarios of
neutrino masses, but a recent work \cite{Lesgourgues:2004ps} has shown
that some approximations in \cite{Hannestad:2002cn} (such as taking
the errors of CMB data given only by cosmic variance) were too
crude. A more conservative estimate \cite{Lesgourgues:2004ps} for
experimental data available approximately at the end of the present
decade is $0.21$ eV for Planck+SDSS, that could be improved to $0.13$
eV with data from CMBpol (a project of a future CMB satellite with
better sensitivity to CMB polarization). As an example, we show in
Fig.\ \ref{fig:Planck} the predicted sensitivity for Planck+SDSS at
2$\sigma$ on the sum of neutrino masses as a function of the assumed
fiducial value.  Note that the possible values of $\sum m_\nu$ are of
course bounded from below: the minimal value corresponds to the limit
in which the lightest neutrino mass goes to zero (different for normal
or inverted hierarchy schemes).

Other cosmological probes of neutrino masses could reach similar or
even better sensitivities in the next future. To probe the mass
distribution of the universe one can use either the weak gravitational
lensing of background galaxies by intervening matter
\cite{Abazajian:2002ck} or the distortions of CMB temperature and
polarization spectra caused by gravitational lensing
\cite{Kaplinghat:2003bh}. These two methods are potentially sensitive
to neutrino masses of the order $0.1$ eV, while the combination of
both could improve it possibly down to the minimum values expected in
the hierarchical neutrino schemes, as recently shown in
\cite{Song:2003gg}.

\section{Conclusions}  

Cosmological data can be used to bound the sum of neutrino masses,
providing information on the absolute neutrino mass scale that is
complementary to terrestrial experiments such as tritium $\beta$ decay
and neutrinoless double $\beta$ decay experiments. We have briefly
described the effects of massive neutrinos on the evolution of the
Universe. The subleading contribution of massive neutrinos to the
cosmological matter content has been analyzed in recent works, in
particular after the the release of the first year WMAP data.  

Current cosmological bounds on the sum of neutrino masses are in the
range $0.42-1.7$ eV (at 95\% CL), which depend both on the included
data and the assumed set of cosmological parameters. These values
prove the region where all neutrino mass states are degenerate, but
future cosmological data will provide sub-eV sensitivities in the
coming years which could test the quasi-degenerate region of neutrino
masses and eventually the minimum values in the hierarchical scenarios,
in particular in the inverted case.

\section*{Acknowledgments}
The author was supported by the Spanish grant BFM2002-00345 and
a Ram\'on y Cajal contract of MEC.

\end{document}